\documentclass[11pt]{article}
\usepackage{graphicx}
\usepackage{palatino}
\usepackage{xspace}
\usepackage{amsmath}
\usepackage{amsfonts}
\usepackage{amssymb}
\usepackage{amsbsy}
\usepackage{lineno}
\usepackage{verbatim}

\newcommand{\BABARPubYear}    {08}

\newcommand{\BABARConfNumber} {015}
\newcommand{\SLACPubNumber} {13339}
\newcommand{\LANLNumber} {0807.4900}

\input pubboard/babarsym
\providecommand{\theLumi}  {\ensuremath{347.3}\xspace}
\providecommand{\thislimitRatBayes}{\ensuremath{0.69\%}\xspace}
\providecommand{\thislimitMoreRatBayes}{\ensuremath{0.85\%}\xspace}
\providecommand{\datEps}{\ensuremath{(-5.5\pm5.8\text{(stat)}^{+0.8}_{-5.5}\text{(syst)})\times10^{-3}}\xspace}

\providecommand{\Km}{\ensuremath{K^{-}}\xspace}
\providecommand{\Kpm}{\ensuremath{K^{\pm}}\xspace}

\providecommand{\piz}{\ensuremath{\pi^0}\xspace}
\providecommand{\pip}{\ensuremath{\pi^+}\xspace}
\providecommand{\pim}{\ensuremath{\pi^-}\xspace}

\providecommand{\Br}{\ensuremath{\mathcal{B}}\xspace}
\providecommand{\fbarn}    {\ensuremath{\text{fb}^{-1}}\xspace}

\providecommand{\epem}  {\ensuremath{e^+e^-}\xspace}

\providecommand{\om}   {\ensuremath{\omega}\xspace}

\providecommand{\ta}   {\ensuremath{\tau}\xspace}

\providecommand{\tm}   {\ensuremath{\tau^-}\xspace}

\providecommand{\nut}  {\ensuremath{\nu_{\tau}}\xspace}
\providecommand{\qqbar}{\ensuremath{q\bar{q}}\xspace}

\providecommand{\tauom}  {\ensuremath{\tm\to\omega\pim\nut}\xspace}

\providecommand{\taub}  {\ensuremath{\tm\to b_{1}^-\nut}\xspace}
\providecommand{\bom}   {\ensuremath{b_1^-\to\omega\pim}\xspace}

\providecommand{\Nratio}{\ensuremath{N^{\omega\pi}_{\text{(non-vector current)}}\,
    /\,N^{\omega\pi}_{\text{(vector current)}}}\xspace} 

\providecommand{\ompppz}{\ensuremath{\om\to\pip\pim\piz}\xspace}
\providecommand{\pppz}{\ensuremath{\pip\pim\piz}\xspace}
\providecommand{\ppppz}{\ensuremath{2\pim\pip\piz}\xspace}
\providecommand{\tppppz}{\ensuremath{\tm\to2\pim\pip\piz\nut}\xspace}
\providecommand{\mtpz}{\ensuremath{m(\ppppz)}\xspace}
\providecommand{\mopz}{\ensuremath{m(\pppz)}\xspace}
\providecommand{\epsil}{\ensuremath{\epsilon}\xspace}
\providecommand{\mt}{\ensuremath{m_\tau}\xspace}
\providecommand{\thet}{\ensuremath{\theta_{\om\pi}}\xspace}
\providecommand{\costhet}{\ensuremath{\cos\thet}\xspace}
\providecommand{\GeVc}{\ensuremath{\mbox{GeV/}c}\xspace}

\providecommand{\MeVcc}{\ensuremath{\mbox{MeV/}c^2}\xspace}

\providecommand{\tauomppz}{\ensuremath{\tm\to\om\pim\piz\nut}\xspace}

\long\def\inst#1{\par\nobreak\kern 4pt\nobreak
    {\it #1}\par\vskip 10pt plus 3pt minus 3pt}

\begin{document}
{\pagestyle{empty}

\begin{flushright}
\babar-CONF-\BABARPubYear/\BABARConfNumber \\
%\babar-PUB-\BABARPubYear/\BABARPubNumber \\
SLAC-PUB-\SLACPubNumber \\
arXiv:\LANLNumber \\
July 2008 \\
%\end{flushright}

%\vspace{.2in}{\bf BAD 2029 v11}\\
\end{flushright}

\par\vskip 5cm

% Title of the paper
\begin{center}
\Large \bf
Search for Second-Class Currents in {\boldmath$\tauom$}
\end{center}
\bigskip

\begin{center}
\large The \babar\ Collaboration\\
\mbox{ }\\
%\today
\end{center}
\bigskip \bigskip

% Abstract
\begin{center}
\large \bf Abstract
\end{center}
We report on an analysis of $\tau^-$ decaying into $\omega\pi^-\nu_\tau$ with 
$\omega\to\pi^+\pi^-\pi^0$ using data containing nearly 320 million tau pairs 
collected with
the BABAR detector at the PEP-II asymmetric energy 
$B$-Factory. We find no evidence 
for second-class currents and 
set an upper limit 
at $0.69\%$ at 
a 90\% confidence level
for the ratio of second- to first-class currents.

\vfill
\begin{center}

Submitted to the 34$^{\rm th}$ International Conference on High-Energy Physics, ICHEP 08,\\
30 July---5 August 2008, Philadelphia, Pennsylvania.

\end{center}

\vspace{1.0cm}
\begin{center}
{\em Stanford Linear Accelerator Center, Stanford University, 
Stanford, CA 94309} \\ \vspace{0.1cm}\hrule\vspace{0.1cm}
Work supported in part by Department of Energy contract DE-AC02-76SF00515.
\end{center}

\newpage
} % end of pagestyle{empty}

% Input author list file
%
%author list removed temporarily to save trees 7/9/04 RNC
%
\begin{center}
\small

The \babar\ Collaboration,
\bigskip

%% author list as of 02-Jul-2008 (523 authors)
%
B.~Aubert,
M.~Bona,
Y.~Karyotakis,
J.~P.~Lees,
V.~Poireau,
E.~Prencipe,
X.~Prudent,
V.~Tisserand
\inst{Laboratoire de Physique des Particules, IN2P3/CNRS et Universit\'e de Savoie, F-74941 Annecy-Le-Vieux, France }
J.~Garra~Tico,
E.~Grauges
\inst{Universitat de Barcelona, Facultat de Fisica, Departament ECM, E-08028 Barcelona, Spain }
% ...

\end{center}\newpage

%\linenumbers
%%%%%%%%%%%%%%%%%%%%%%%%%%%%%%%%%%%%%%%%%%%%%%%%%%%%%%%%%%%%%%%%%%%%%%%%%
\section{INTRODUCTION}
\label{sec:Introduction}

Weak currents can be classified as either 
first- or second-class
depending on the $J^{PG}$ of the decay current \cite{scc},
where $G$-parity
is a combination of charge conjugation and an isospin 
rotation, $\hat{G}=\hat{C}e^{i\pi\hat{I}_2}$, and is 
a multiplicative quantum number.  In the 
Standard Model, first-class currents (FCC), where $PG(-1)^J=+1$ 
($J^{PG} = 0^{++}, 0^{--}, 1^{+-}, 1^{-+},\ldots$), are expected to
dominate decays while second-class currents (SCC), where $PG(-1)^J=-1$
($J^{PG} = 0^{+-}, 0^{-+}, 1^{++}, 1^{--},\ldots$),
are expected to be small and to vanish in the limit of
perfect isospin symmetry.
An example of such a decay is 
$\tm\to\omega\pim\nut$\footnote{Charge-conjugate 
reactions are implied throughout this paper.}, which 
is expected to proceed through FCC
mediated by the $\rho$ resonance.
This decay may also potentially proceed through SCC,
such as 
$b_1$(1235)~\cite{leroy} with $\taub\to\om\pim\nut$,
producing final state particles with 
$J^{PG}=1^{++}$ and $0^{-+}$.

\begin{figure}[ht]
\begin{center}
  \includegraphics[width=0.4\textwidth]{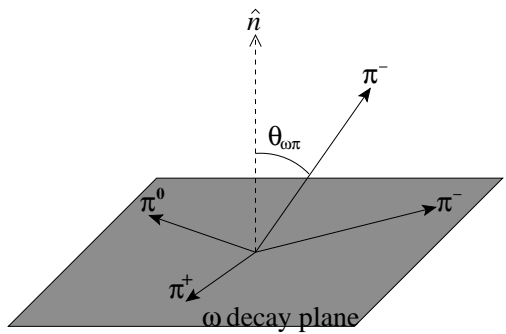}
\end{center}
\caption{Illustration of the angle $\thet$: 
the angle between the normal to the \om decay plane and the direction 
of the remaining $\pi$ in the \om\ rest frame.}
\label{fig:costh}
\end{figure}

Since the decay \bom occurs through S- and D-waves, as compared
to a P-wave for FCC,
different polarizations of \om spin result
in different angular distributions of the final state particles.
The expected distributions of \costhet for all possible spin-parity
states of the final state particles are listed in Table~\ref{table:sccshape}, 
where $\thet$ is the angle between the normal to the \om decay
plane and the direction of the remaining $\pi$ in the \om rest frame,
as shown in Figure~\ref{fig:costh}.
The existing measurement of the
angular distribution of \tauom is
consistent with having only P-wave contribution, and the present limit 
is 5.4\% for the ratio of SCC to FCC contributions,
\Nratio,
at 90\% confidence level \cite{cleoscc}. 
This paper presents a search for SCC in \tauom\ decays with \ompppz
by studying the angular distributions of final state particles.

\begin{table}[h!]
\caption{Expected angular distributions, $F_L(\costhet)$, for possible
spin-parity states in the decay of \tauom. L is the orbital angular
momentum.}
\begin{center}
\begin{tabular}{|c|c|c|c|}
\hline
$J^{P}$ & L  & $F^{FCC}$ & $F^{SCC}_L(\costhet)$ \\
\hline
$1^-$ & 1 &  $F^{FCC}\propto (1-\mbox{cos}^2\thet)$ & \\
$1^+$ & 0 &  & $F_0^{SCC}\propto 1$\\
$1^+$ & 2 &  & $F_2^{SCC}\propto (1+3\mbox{ cos}^2\thet)$\\
$0^-$ & 1 &  & $F_1^{SCC}\propto \mbox{cos}^2\thet$\\
\hline
\end{tabular}

\end{center}
\label{table:sccshape}
\end{table}

%%%%%%%%%%%%%%%%%%%%%%%%%%%%%%%%%%%%%%%%%%%%%%%%%%%%%%%%%%%%%%%%%%%%%%%%%
\section{THE \babar\ DETECTOR AND DATASET}
\label{sec:babar}

This analysis is based on data recorded 
by the \babar\ detector \cite{babar} at the \pep2\ asymmetric-energy 
\epem\ storage rings operated at the Stanford Linear Accelerator Center. 
The data sample consists of \theLumi\ \fbarn recorded at
the center-of-mass energy of $10.58\gev$.
With a cross section for \ta\ pairs
of $\sigma_{\tau\tau} = (0.919\pm0.003)$ nb \cite{tauxsect},
this data sample contains nearly 320 million pairs of tau decays.
 
The \babar\ detector is described in detail in Ref.~\cite{babar}.
Charged-particle momenta are measured with a 5-layer
double-sided silicon vertex tracker (SVT) and a 40-layer drift chamber (DCH)
inside a 1.5-T superconducting solenoidal magnet.
An electromagnetic calorimeter (EMC) consisting of 6580 CsI(Tl) 
crystals is used to identify electrons and photons.
A ring-imaging Cherenkov detector is used to identify
charged hadrons, in combination with ionization energy loss measurements 
($dE/dx$) in the SVT and the DCH.
Muons are identified 
by an instrumented magnetic-flux return (IFR).

Monte Carlo (MC) simulations are used to estimate 
the signal efficiencies and background contamination.
The production of \ta\ pairs is
simulated with the {\tt KK2f} 
generator \cite{kk2f}, and  the decays of the 
\ta\ lepton are modeled with {\tt Tauola} \cite{tauola}.
Continuum \qqbar\ events are simulated using {\tt JETSET} \cite{jetset}.
Final state radiative effects are simulated for all 
decays using {\tt Photos} \cite{photos}.
The detector response is simulated 
with {\tt GEANT4} \cite{geant}, and the simulated events are then 
reconstructed in the same manner as data. 

%%%%%%%%%%%%%%%%%%%%%%%%%%%%%%%%%%%%%%%%%%%%%%%%%%%%%%%%%%%%%%%%%%%%%%%%%
\section{ANALYSIS METHOD}
\label{sec:Analysis}

%% -> 1. event selection
Since \ta\ pairs are produced back-to-back in 
the \epem\ center-of-mass frame, each event is divided into two
hemispheres according to the thrust axis \cite{thrust}, calculated 
using all reconstructed charged particle tracks.
Candidate events in this analysis are required to 
have a ``1-3 topology,'' where one track is in one hemisphere 
(tag hemisphere) and three tracks are in the other hemisphere 
(signal hemisphere). 
Events with four well-reconstructed tracks and zero net charge 
are selected.
The polar angles, in the laboratory frame, of all four tracks 
and the neutrals used in \piz reconstruction
are required to be within the calorimeter acceptance 
range. Events are rejected if the invariant mass of 
pairs of oppositely charged tracks, assuming electron mass hypotheses,
is less than 90 \MeVcc, as these tracks are likely to be 
from photon conversions in the detector material. 

The charged particle found in the tag hemisphere 
must be either an electron or a muon 
candidate. Electrons are identified using the ratio of 
calorimeter energy to track momentum ($E/p$), the shape of 
the shower in the calorimeter, and $dE/dx$.
Muons are identified by hits in the IFR and small energy 
deposits in the calorimeter consistent with expectation
for a minimum-ionizing particle. 
Muons with momentum less than 0.5 \GeVc cannot be identified 
in this manner
as they do not penetrate far enough into the IFR.
Charged particles found in the signal hemisphere must 
be identified as pion candidates using $dE/dx$.
The \piz candidates are reconstructed from 
two separate EMC clusters with energies above 100 MeV that are not
not associated with charged tracks and are required to have
invariant masses % of \piz candidates is
between 100 and  160 \MeVcc.
Events are required to have a single \piz in the signal hemisphere.
The \ta candidates are reconstructed in the signal hemisphere
using the three tracks and the \piz candidate, and the invariant 
mass of the \ta candidate, $\mtpz$ is required to be less than 
the nominal mass of the \ta lepton, $\mt = 1.777\gevcc$ \cite{pdg}.
After the event selection process, from the MC it is found that 
14\% of the events remaining 
are \ta-pair events that do not contain a \tppppz decay, 
and 1.3\% are \epem\to\qqbar events.

\begin{figure}[tb]
\begin{center}
  \includegraphics[width=0.8\textwidth]{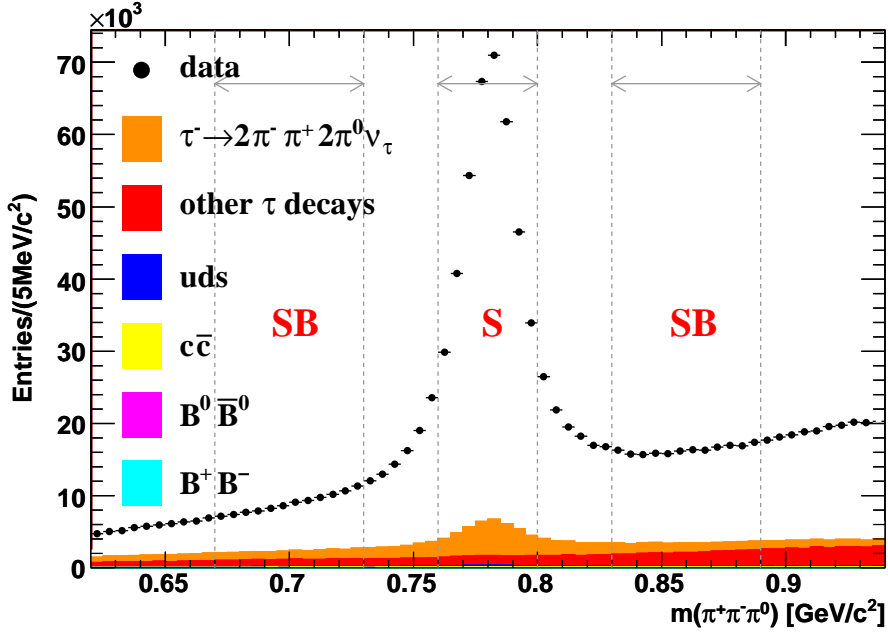}
\end{center}
\caption{\om candidate mass spectra for selected events in data and 
  expected Monte Carlo background (colored histograms).  The background
  histograms do not include the non-resonant \tppppz decays.
  The signal (S) and sideband (SB) regions are indicated in the figure.}
\label{fig:ommass}
\end{figure}

%% -> 2. analysis method (omega definition)
For each selected event with \mtpz $<$ \mt 
two \om\ candidates are reconstructed from \pppz\ combinations.
The mass of the \om\ candidates, \mopz, is required to be between
670 \MeVcc and 890 \MeVcc; within this range, the signal
region is defined between 760 \MeVcc and 800 \MeVcc with 
mass regions of width 60 \MeVcc on each side of the peak used
as sideband regions for background studies, as shown in 
Figure~\ref{fig:ommass}.
For each \om\ candidate in the signal region, the angle $\thet$ 
is calculated and is used in the SCC measurement, 
after background subtraction.

%% 3. background
There are three background types to be considered 
in this analysis. 
The first type is combinatoric background, which
is expected to have an angular distribution that is 
independent of \mopz, and is thus
subtracted from the signal region using the sideband regions. 
The number of combinatoric events lying within the signal region 
is obtained by fitting the \mopz spectrum with a smeared relativistic 
Breit-Wigner for the \om resonance and a polynomial for the 
combinatoric background.  The polynomial is integrated over the 
signal region to find the number of continuum events in the 
signal region. 
The second type of background comes from \epem\to\qqbar\ events that 
contain $\om\to\pip\pim\piz$ decays. 
While the event selection
process significantly reduces the number of \qqbar\ events, 
approximately 0.3\% of the 
events in the signal region are expected to be of \qqbar origin.
This type of background is studied using 
events with $m(\ppppz)$ 
well above the \ta mass ($>2.1\gevcc$).
In this region, where all events are considered to be of \qqbar origin,
a comparison of the numbers of events in MC and data is used to 
obtain a scaling factor for the \qqbar background events. 

\begin{comment}
The angular 
distribution obtained for the \epem\to\qqbar\ events containing \om 
decays is shown in Figure~\ref{fig:bkg costh}(b).

\begin{figure}[tb]
\begin{center}
  \begin{tabular}{cc}
    \includegraphics[width=.5\textwidth]{eps/combinatoric.eps} & 
    \includegraphics[width=.5\textwidth]{eps/qqbarbkg.eps}\\
    (a) & (b)\\
  \end{tabular}
\end{center}
\caption{Background \costhet distributions in the signal region for 
  (a) combinatoric background, (b) resonant \epem\to\qqbar background}
\label{fig:bkg costh}
\end{figure}
\end{comment}

\begin{figure}[hbt]
\begin{center}
  \begin{tabular}{cc}
    \includegraphics[width=.5\textwidth]{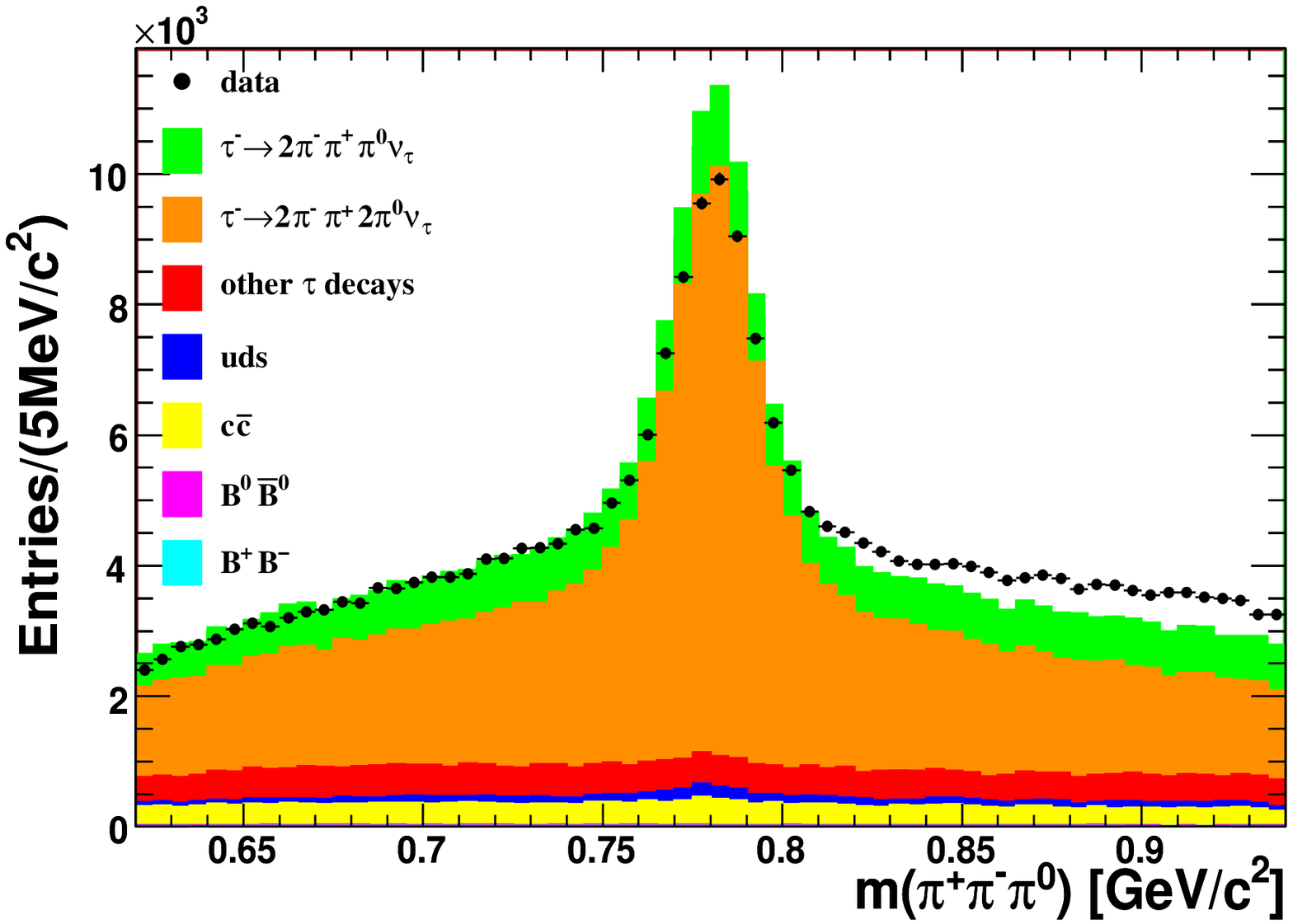} & 
    \includegraphics[width=.5\textwidth]{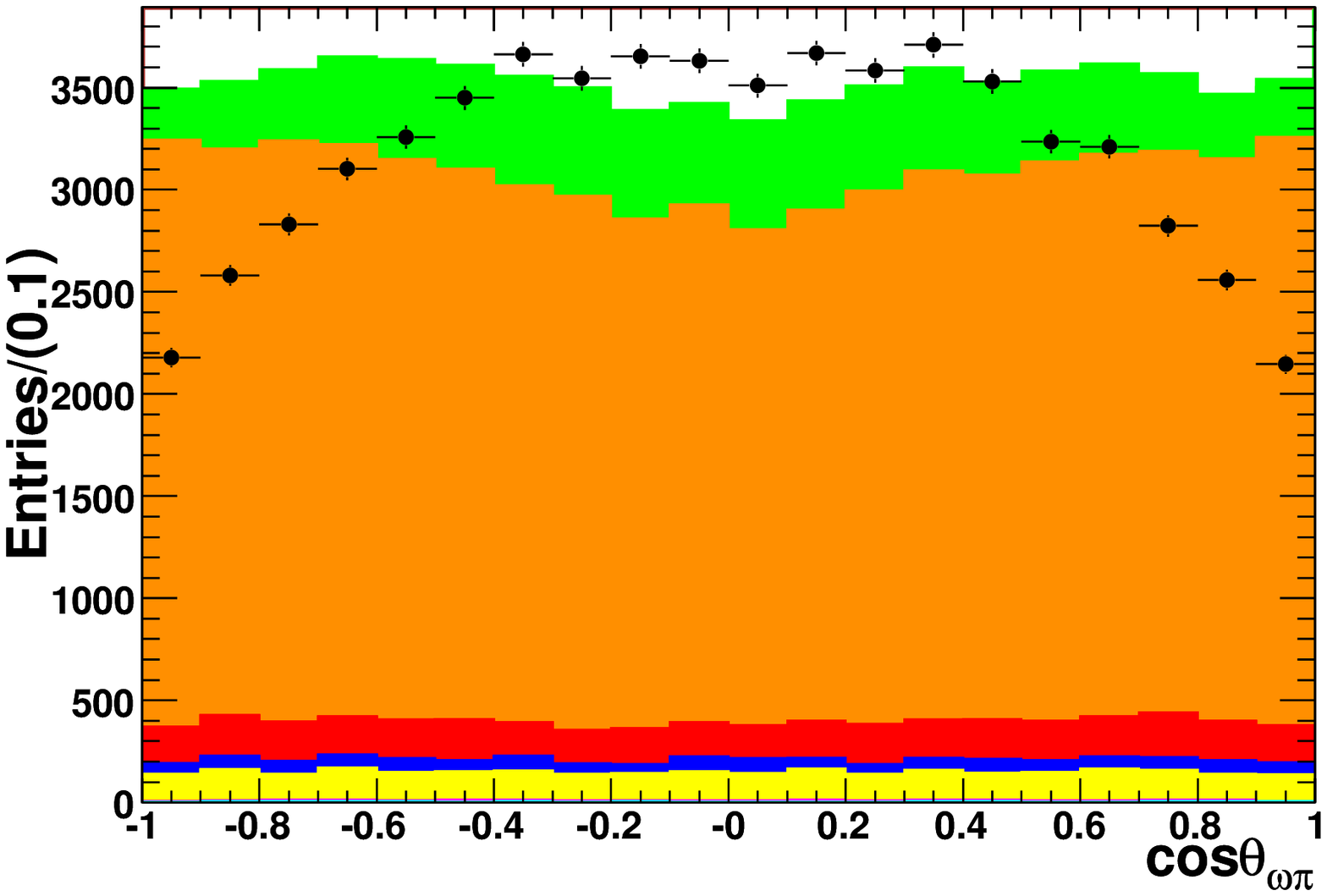}\\
    (a) & (b)\\
  \end{tabular}
  \includegraphics[width=.5\textwidth]{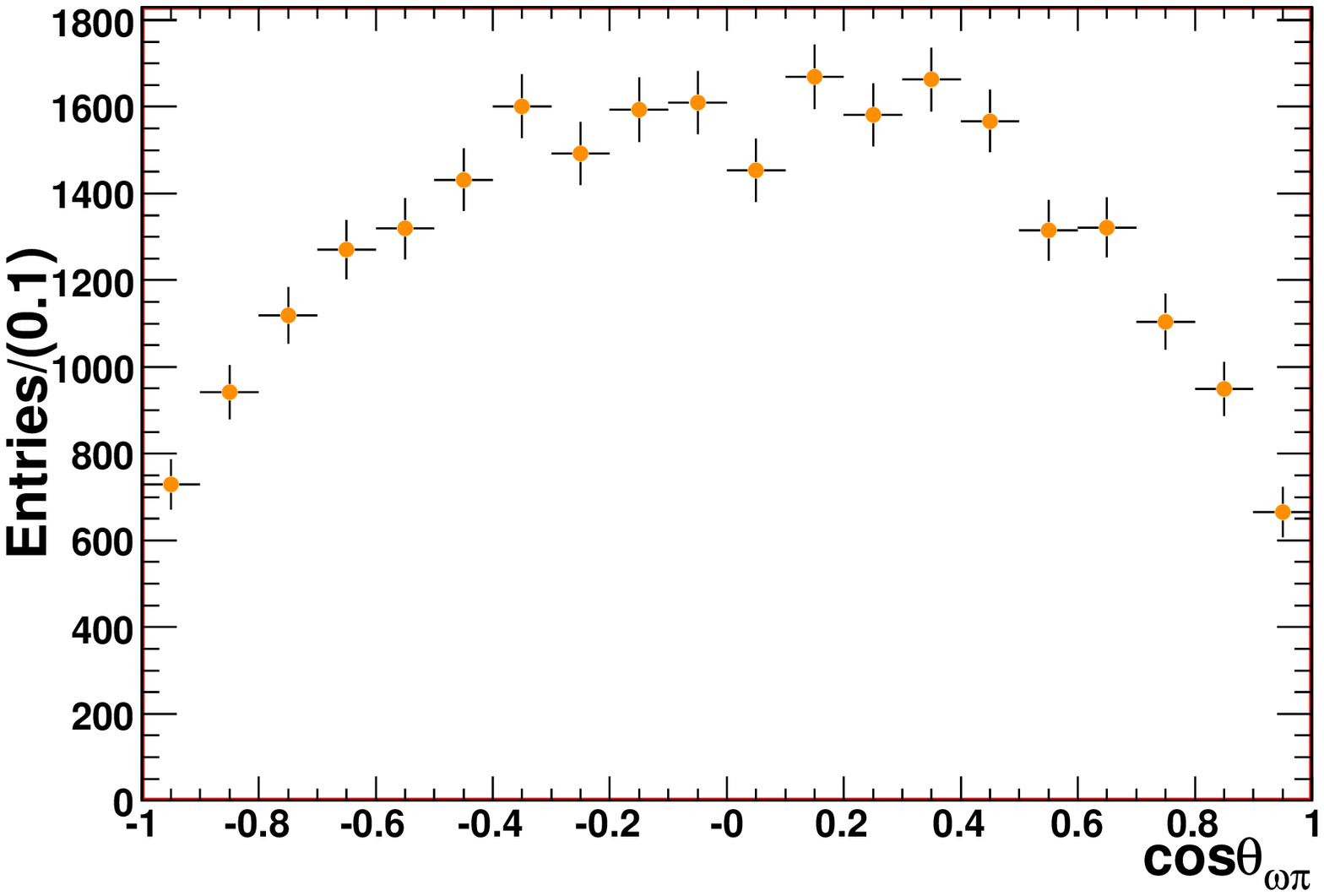}\\
  (c)
\end{center}
\caption{(a) $m(\pip\pim\piz)$ and (b) \costhet distributions 
when requiring an additional \piz in the signal hemisphere, before
background subtraction. These are used to correct the 
$\tm\to\om\pim\piz\nut$ MC.
(c)~The \costhet distribution obtained from data after subtracting
background.}
\label{fig:george}
\end{figure}

After subtracting combinatoric and \qqbar background events, approximately 
4.6\% of the remaining
\om candidates in the signal region are expected to be background
events from non-signal \ta decays. 
The dominant of these, comprising 99\% of these background events, 
is \tauomppz, where one \piz has not been reconstructed.
The decay \tauomppz has not been well measured and is 
incorrectly modeled in the MC. 
Both the branching fraction and decay angular distribution
need to be corrected, as shown in Figures 
\ref{fig:george}(a) and \ref{fig:george}(b).
To correct for the differences between data and MC,
events with an additional \piz candidate in the signal hemisphere
are selected, using the same cuts discussed above.
Using these events, the MC branching fraction of \tauomppz is 
corrected by comparing the numbers of fitted \om candidates in 
data and MC. The fit function used for this is a smeared 
relativistic Breit-Wigner with a polynomial background.
The branching fraction obtained using this correction
technique is found to be consistent 
with existing measurements \cite{pdg}.
To correct the angular distribution of \tauomppz in the signal region,
backgrounds, consisting of combinatorics, \qqbar events and \tauom decays,
are subtracted from the two \piz data sample, and
the remaining \costhet distribution, 
shown in Figure \ref{fig:george}(c), is used to correct
the \tauomppz distribution in the MC.

\begin{figure}[hbt]
\begin{center}
  \includegraphics[width=.45\textwidth]{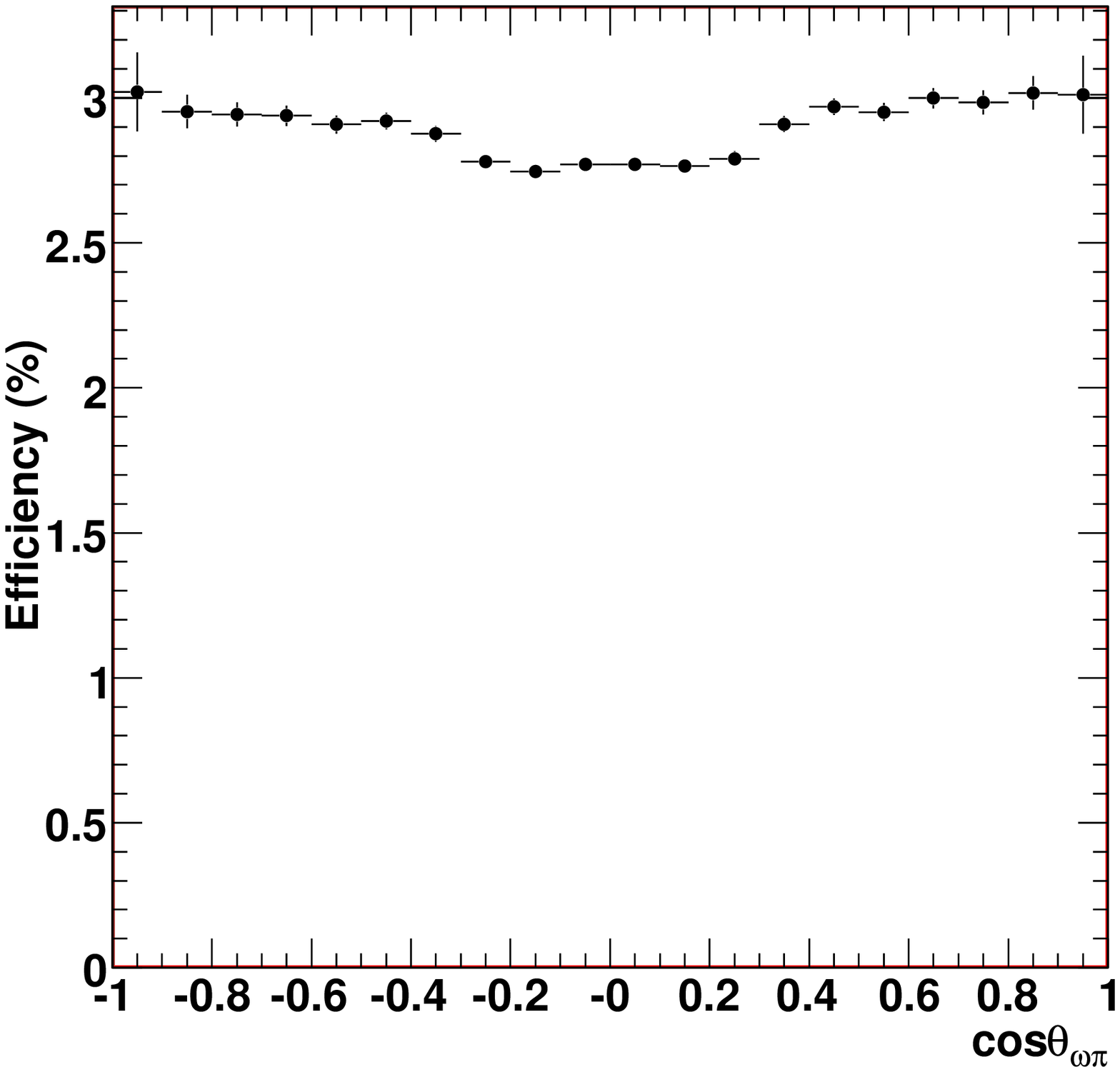}\\
\end{center}
\caption{Efficiency as a function of \costhet obtained from 
\tppppz MC.}
\label{fig:efficiency}
\end{figure}

To account for any variation in efficiency as a function of \costhet, 
the generated and reconstructed MC \costhet distributions 
of \tauom in the signal region are compared.
The ratio of the two 
distributions, shown in Figure \ref{fig:efficiency}, 
is used as an efficiency function to correct the background 
subtracted data.

%%%%%%%%%%%%%%%%%%%%%%%%%%%%%%%%%%%%%%%%%%%%%%%%%%%%%%%%%%%%%%%%%%%%%%%%%
\section{PHYSICS RESULTS}
\label{sec:Physics}

%% -> 2.1: analysis method = fitting
After subtracting background events and applying efficiency corrections, 
a binned fit to the remaining \costhet\ distribution is carried out using 
\begin{equation}
  F(\costhet)\,
  =N\times[\epsilon F_{0}^{SCC}(\costhet)+(1-\epsilon) F^{FCC}(\costhet)],
\label{eq:onescc}
\end{equation}
where $N$ is a normalization factor, the parameter $\epsilon$ is the fraction 
of \tauom decays that proceed through SCC,
and $F^{FCC}$ and $F_0^{SCC}$ 
are normalized angular functions 
described in Table \ref{table:sccshape}. The parameter \epsil is related to 
\Nratio by the equation $\eps/(1-\eps)=\Nratio$.
In Eq.\ref{eq:onescc}, only $F_0^{SCC}$ is used for the function
describing the SCC contribution since the shape of this
function gives the most conservative (largest) estimate of \epsil.

This method is tested by  adding various amounts of $S$-wave
decays to the standard MC and fitting for the levels of SCC.
As the results of this test indicate, as shown in Table \ref{table:scctest},
in all cases the measured fractions of SCC, \epsil, are
consistent with the fractions added to the MC. 
The errors listed in Table \ref{table:scctest} are not necessarily 
indicative of the expected uncertainties in the data; they do not contain
systematic uncertainties, and statistical correlations exist among the
MC samples used in the studies.

\begin{table}[ht]
\caption{Test of SCC measurements in MC with various amounts
of $S$-wave decays added.}
\begin{center}
\begin{tabular}{c|c}
\hline
Fraction of $L=0$ \tauom decays added & $\epsilon$\\
\hline
none  & $(0.11\pm 0.17)\%$ \\
1\%   & $(1.10\pm 0.17)\%$ \\
2\%   & $(2.09\pm 0.17)\%$ \\
\hline
\end{tabular}
\end{center}
\label{table:scctest}
\end{table}

% Systematics...
%%%%%%%%%%%%%%%%%%%%%%%%%%%%%%%%%%%%%%%%%%%%%%%%%%%%%%%%%%%%%
The largest contributions to systematic uncertainties on \epsil are
scaling and modeling of the MC background. The correction applied 
to the branching fraction of \tauomppz has an error associated with 
it, determined by the available statistics. The correction factor 
is adjusted by $\pm1\sigma$ to obtain
the uncertainty in \epsil while the errors associated with 
correcting the angular distribution are folded into the statistical
uncertainty. In addition, there are \ta decays that may be present in 
the final event sample but which are not simulated in the MC. The largest 
of these are expected to be $\tm\to\omega K^-\nut$,
$\tm\to\omega\pim2\piz\nut$ and $\tm\to\omega 2\pim\pip\nut$ decays, 
which when combined can add up to 0.2\% of the final event sample.
Since the effect that these decays have on the angular distribution
is unknown, the extreme cases are taken to obtain the uncertainty.
These cases correspond to these decays having either entirely 
$1-\cos^2\thet$ or entirely $\cos^2\thet$ distributions.
The scaling of \qqbar events can also affect the measurement 
of \epsil, and the uncertainty is obtained by adjusting the scaling
factor by $\pm1\sigma$.
These systematic uncertainties are summarized in Table \ref{table:syst}.

\begin{table}[ht]
\caption{Summary of systematic uncertainties on \epsil}
\begin{center}
\begin{tabular}{c|c}
\hline
Source & Uncertainty ($\sigma_\epsil$) \\
\hline
\Br(\tauomppz) & $\pm0.0007$\\
un-simulated \ta decays & $_{-0.0055}^{+0.0000}$ \\
\qqbar scaling & $\pm0.0001$\\
\hline
Total & $_{-0.0055}^{+0.0008}$ \\
\hline
\end{tabular}
\end{center}
\label{table:syst}
\end{table}

To estimate the sensitivity of this analysis without 
the effect of statistical correlations in the MC samples used,
a toy MC study is conducted.
In this study, angular distributions are
generated for the signal and sideband regions to simulate the statistics 
available in the data and various MC samples used in the analysis.  
After subtracting background samples from the toy data, the 
angular distribution is corrected for efficiency and fitted
using Eq.\ref{eq:onescc}.
The statistical uncertainty on \epsil obtained from the fit is 
$0.0063$, which combined with the systematic uncertainties
leads to an estimated uncertainty of 
$\sigma_\epsil = _{-0.0084}^{+0.0064}$.

\begin{figure}[hbt]
\begin{center}
  \includegraphics[width=0.7\textwidth]{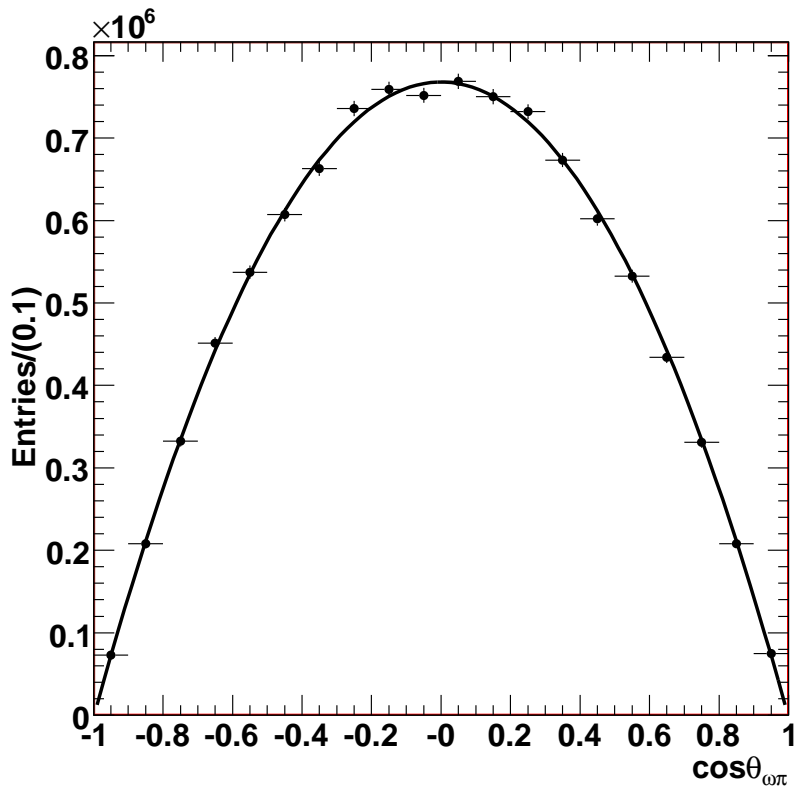}
\end{center}
\caption{
  The fitted \costhet distribution for the data. The fitted curve is
  described in the text.}
\label{fig:dataCosth}
\end{figure}

%%%%% RESULTS %%%%%%%
With the MC studies completed, the angular distribution
in the data is obtained by subtracting estimated background
events as described above.
The remaining distribution is corrected for efficiency and fitted using
Eq.~\ref{eq:onescc} as shown in Figure \ref{fig:dataCosth}. 
The fit has $\chi^2/dof =$ 15.4/18, and the fitted value of
\epsil in the data is \datEps, which is consistent with no SCC 
contribution to \tauom decays.
For the upper limit on \Nratio, a Bayesian approach~\cite{bayesian}
is used as negative values of \epsil are non-physical. 
Using only the positive portion 
of the probability distribution for the value of the SCC contribution, 
$\epsilon_{true}$, 
the distribution for $\epsilon_{true}$ is a
bifurcated Gaussian with mean $\epsilon=5.5\times10^{-3}$ and errors 
$\sigma_\epsilon=_{-0.0084}^{+0.0064}$.
The limits obtained from this method 
are \thislimitRatBayes at 90\% C.L. 
and \thislimitMoreRatBayes at 95\% C.L.

%%%%%%%%%%%%%%%%%%%%%%%%%%%%%%%%%%%%%%%%%%%%%%%%%%%%%%%%%%%%%%%%%%%%%%%%%
\section{SUMMARY}
\label{sec:Summary}

A search for second-class currents in the decay \tauom is 
conducted with the \babar\ detector. 
No evidence for second-class currents is observed, 
and a 90\% confidence level Bayesian upper limit for \Nratio\ is
set at \thislimitRatBayes. 
This limit is an order of magnitude lower than
the limit set by the CLEO collaboration \cite{cleoscc}.

%%%%%%%%%%%%%%%%%%%%%%%%%%%%%%%%%%%%%%%%%%%%%%%%%%%%%%%%%%%%%%%%%%%%%%%%%
\section{ACKNOWLEDGMENTS}
\label{sec:Acknowledgments}
We are grateful for the 
extraordinary contributions of our \pep2\ colleagues in
achieving the excellent luminosity and machine conditions
that have made this work possible.
The success of this project also relies critically on the 
expertise and dedication of the computing organizations that 
support \babar.
The collaborating institutions wish to thank 
SLAC for its support and the kind hospitality extended to them. 
This work is supported by the
US Department of Energy
and National Science Foundation, the
Natural Sciences and Engineering Research Council (Canada),
the Commissariat \`a l'Energie Atomique and
Institut National de Physique Nucl\'eaire et de Physique des Particules
(France), the
Bundesministerium f\"ur Bildung und Forschung and
Deutsche Forschungsgemeinschaft
(Germany), the
Istituto Nazionale di Fisica Nucleare (Italy),
the Foundation for Fundamental Research on Matter (The Netherlands),
the Research Council of Norway, the
Ministry of Education and Science of the Russian Federation, 
Ministerio de Educaci\'on y Ciencia (Spain), and the
Science and Technology Facilities Council (United Kingdom).
Individuals have received support from 
the Marie-Curie IEF program (European Union) and
the A. P. Sloan Foundation.

%%%%%%%%%%%%%%%%%%%%%%%%%%%%%%%%%%%%%%%%%%%%%%%%%%%%%%%%%%%%%%%%%%%%%%%%%

\end{document}